# A Non-stochastic Optimization Algorithm for Neural-network Quantum States

Xiang Li[1], Jia-Cheng Huang[1], Guang-Ze Zhang[1], Hao-En Li[1], Chang-Su Cao[1,2], Dingshun Lv[2], Han-Shi Hu[1,*]

[1]*Department of Chemistry and Engineering Research Center of Advanced Rare-Earth Materials of Ministry of Education, Tsinghua University, Beijing 100084, China*

[2]*ByteDance Research, Zhonghang Plaza, No. 43, North 3rd Ring West Road, Haidian District, Beijing, 100089, China*

(Dated: July 2023)

**ABSTRACT:** Neural-network quantum states (NQS) employ artificial neural networks to encode many-body wave functions in second quantization through variational Monte Carlo (VMC). They have recently been applied to accurately describe electronic wave functions of molecules and have shown the challenges in efficiency comparing with traditional quantum chemistry methods. Here we introduce a general non-stochastic optimization algorithm for NQS in chemical systems, which deterministically generates a selected set of important configurations simultaneously with energy evaluation of NQS. This method bypasses the need for Markov-chain Monte Carlo within the VMC framework, thereby accelerating the entire optimization process. Furthermore, this newly-developed non-stochastic optimization algorithm for NQS offers comparable or superior accuracy compared to its stochastic counterpart and ensures more stable convergence. The application of this model to test molecules exhibiting strong electron correlations provides further insight into the performance of NQS in chemical systems and opens avenues for future enhancements.

## 1. INTRODUCTION

Various methods searching for solutions of the many-electron Schrödinger equation have been developed with a trade-off between efficiency and accuracy, among which the full configuration interaction (FCI) method is exact as it considers all electron configurations but is limited to small molecules due to its high computational cost. Thus, constructing wave function ansatzes to reduce the exponential complexity of FCI one down to its most essential features has been under steady progress. One prime example is the selected configuration interaction (SCI) method[1-13], which capitalizes on the paucity of configurations that contribute appreciably to the exact one and tackles large configuration space by iteratively augmenting a selected small set of important configurations.

This strategy for tackling the exponential scaling of the FCI wave function shares similarities with dimensional reduction and feature extraction in data science[14], suggesting the potential of leveraging the power of artificial neural networks (ANN). In 2017, Carleo and Troyer first introduced an ANN model based on the restricted Boltzmann machine (RBM), referred to as neural-network quantum states (NQS), to encode the whole many-body wave function with high precision in discrete spin lattice systems through variational Monte Carlo (VMC)[15]. This is an ab initio and variational approach that requires no pre-existing data nor prior knowledge of the exact wave function to train the neural network and has no fundamental limits to its accuracy. The success of RBM-based NQS has also been extended to approximating the fermionic wave function of chemical molecules in second-quantized form[16, 17], achieving remarkable accuracy on test molecules. Additionally, the applicability of this innovative idea in continuous space[18, 19] has been explored, and various other recent applications have been summarized[20].

However, there are still challenges when applying the NQS to large chemical systems, primarily due to the inherent MC sampling procedures. The first challenge is the inefficiency of MC sampling in dealing with the peculiar structure of molecular ground-state wave function, which typically exhibits sharp peaks around the Hartree-Fock state and neighboring excited states[16]. A recent proposal has attempted to circumvent this limitation, but it is only applicable with the autoregressive models[21-23]. Another challenge is associated with the statistical error arising from MC sampling, which is proportional to the inverse square root of the number of sampling points used[24, 25]. This issue becomes particularly problematic when calculating small energy differences.

To address these two challenges with MC sampling, we introduce in this article the incorporation of SCI methods into a non-stochastic optimization algorithm for NQS. In this approach, a set of important configurations identified through the modulus of NQS is deterministically selected simultaneously with the energy evaluation of NQS. This non-stochastic configuration selection substitutes the MC configuration sampling required in the stochastic NQS optimization and circumvents the significant computational bottleneck caused by MC sampling as the selection is integrated into the inherent energy evaluation with little extra cost. Furthermore, the stochastic noise introduced by MC sampling is naturally eliminated.

The rest of this paper is organized as follows. In section 2, we provide a brief overview of RBM-based NQS and the unsupervised, variational, and stochastic optimization algorithm based on MC sampling. Section 3 elaborates on the process of deterministically selecting desired configurations during energy evaluation and highlights the distinctions between our new non-stochastic optimization algorithm and the standard MC optimization algorithm. Next, in section 4, we compare the accuracy and efficiency of our non-stochastic algorithm with the stochastic one and benchmark the accuracy against FCI results. The computational details can be referenced in the Supporting Information. Finally, in section 5, we present the conclusions.

## 2. NEURAL-NETWORK QUANTUM STATES

**2.1. Representation of Many-electron Wave Functions.** The fundamental concept behind neural-network quantum states is to represent the many-body wave function in a second-quantized form using one flexible neural network. When applying NQS to quantum chemistry, we consider the many-electron wave function in the following form:

$$|\Psi\rangle = \sum_k \psi_k |D_k\rangle, \tag{1}$$

where $|D_k\rangle = |\sigma_1^k, \sigma_2^k, \dots, \sigma_M^k\rangle$ represents an electron configuration in Slater determinant form. Here, $\sigma_i^k \in \{0,1\}$ denotes the occupation number of each electronic spin orbital, and $M$ is the total number of electronic spin orbitals. The coefficient $\psi_k$ of configuration $|D_k\rangle$ is crucial in determining the exact wave function and can be represented by a neural network.

In this context, we specialize our discussion to the complex-valued restricted Boltzmann machine as depicted in Figure 1, which has proven successful in various interacting quantum models[15, 26-31] and several chemical molecules[16, 17]. This impressive performance can be partly attributed to the intimate connection between RBM and two-dimensional tensor network[30, 32, 33]. In this work, we describe RBM wave function ansatz using this compact form

$$|\Psi_\theta\rangle = \sum_k \psi_\theta(\sigma_1^k, \sigma_2^k, \ldots, \sigma_M^k)|D_k\rangle, \qquad (2)$$

where the coefficient function $\psi_\theta(\sigma_1^k, \sigma_2^k, \ldots, \sigma_M^k)$ is defined as

$$\psi_\theta(\sigma_1^k, \sigma_2^k, \ldots, \sigma_M^k) = e^{\sum_i^M a_i \sigma_i^k} \prod_j^{\lfloor \alpha \cdot M \rfloor} \left(1 + e^{b_j + \sum_i^M W_{ij} \sigma_i^k}\right). \qquad (3)$$

Here, $\theta$ represents a set of complex-valued network parameters, including the input biases $a_i$ ($i = 1,2,\ldots,M$) as a vector, the hidden biases $b_j$ ($j = 1,2,\ldots,\alpha \cdot M$) as a vector, and the kernel weights $W_{ij}$ as a matrix. The total number of the required parameters can be calculated as $M + \alpha \cdot M + \alpha \cdot M^2$. The hidden unit density $\alpha$ is a constant value and corresponds to the ratio of the number of hidden units in RBM to the number of input units. A larger value of $\alpha$ indicates a stronger expressive power of the RBM ansatz.

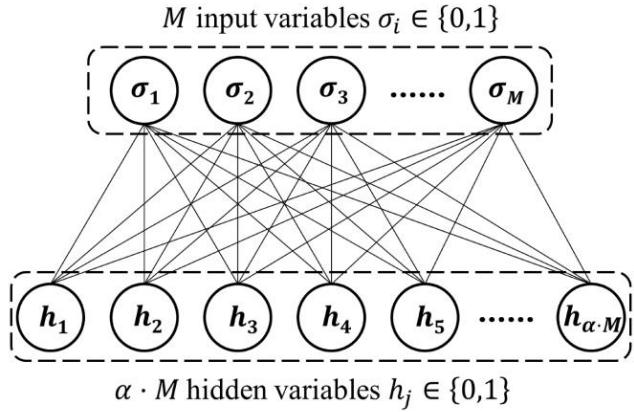

**Figure 1.** Architecture of the RBM-based NQS. The RBM consists of a number of input units which is equal to the number of electronic spin orbitals $M$ in the chemical system. The number of hidden units is determined by the hidden unit density $\alpha$, which is multiplied by the number of input units. The hidden unit density $\alpha$ governs the expressive power of the RBM wave function ansatz.

**2.2. Variational Monte Carlo Optimization.** The NQS wave function ansatz is typically optimized using the variational Monte Carlo approach to approximate the exact ground-state wave function. According to the variational principle, it is achieved by iteratively minimizing the energy expectation value

$$E_\theta = \frac{\langle \Psi_\theta | \hat{H} | \Psi_\theta \rangle}{\langle \Psi_\theta | \Psi_\theta \rangle} = \sum_k P(D_k) E_{loc}(D_k), \tag{4}$$

where we have defined the probability $P(D_k)$ and local energy $E_{loc}(D_k)$ of a given electron configuration as

$$P(D_k) = \frac{|\langle D_k | \Psi_\theta \rangle|^2}{\sum_i |\langle D_i | \Psi_\theta \rangle|^2}, \text{ and} \tag{5}$$

$$E_{loc}(D_k) = \frac{\langle D_k | \hat{H} | \Psi_\theta \rangle}{\langle D_k | \Psi_\theta \rangle} = \sum_\mu \frac{\langle D_\mu | \Psi_\theta \rangle}{\langle D_k | \Psi_\theta \rangle} \langle D_k | \hat{H} | D_\mu \rangle. \tag{6}$$

The MC method is employed to calculate the energy, which averages the local energy over a configuration sample $\mathcal{V}$ drawn from the probability distribution $P(D_k)$. This sample $\mathcal{V}$ consists of $N_{tot}$ configurations which can be generated using the Metropolis-Hastings algorithm[34], and can be represented by $N_{unique}$ unduplicated configurations $\{|D_k\rangle\}$ with their respective repetition counts $N(D_k)$. Thus, the inaccessible accurate probability $P(D_k)$ is estimated as $P(D_k) \approx \frac{N(D_k)}{N_{tot}}$, where $N_{tot} = \sum_{k=1}^{N_{unique}} N(D_k)$. The energy of NQS can then be approximated by

$$E_\theta \approx \sum_{k \in \mathcal{V}} \frac{N(D_k)}{N_{tot}} E_{loc}(D_k). \tag{7}$$

Gradient-based iterative algorithms are subsequently employed to minimize the energy of the NQS with respect to its parameters $\theta$. As the energy $E_\theta$ is a real-valued function with complex-valued parameters, the negative gradient with respect to the conjugate of the neural-network parameters defines the direction of the steepest descent according to Wirtinger calculus[35]. The gradient can be evaluated as

$$g_{\theta_m} = \frac{\partial E_\theta}{\partial \theta_m^*} = \frac{\langle \Psi'_{\theta_m} | \hat{H} | \Psi_\theta \rangle}{\langle \Psi_\theta | \Psi_\theta \rangle} - E_\theta \frac{\langle \Psi'_{\theta_m} | \Psi_\theta \rangle}{\langle \Psi_\theta | \Psi_\theta \rangle}, \tag{8}$$

where $|\Psi'_{\theta_m}\rangle = \sum_k \frac{\partial \psi_\theta(\sigma_1^k, \sigma_2^k, \ldots, \sigma_M^k)}{\partial \theta_m} |D_k\rangle$. This complicated formula can be approximated using the configuration sample $\mathcal{V}$ introduced above, yielding,

$$g_{\theta_m} \approx \sum_{k \in \mathcal{V}} P(D_k) O^*_{k\theta_m} [E_{loc}(D_k) - E_\theta], \tag{9}$$

where $O_{k\theta_m} = \frac{\partial \psi_\theta(\sigma_1^k, \sigma_2^k, ..., \sigma_M^k)}{\partial \theta_m}$.

Finally, the stochastic reconfiguration (SR) scheme[36], a second-order method which can be viewed as an approximate imaginary time evolution in the variational subspace, is employed to ensure a faster and more robust convergence compared to the first-order stochastic gradient descent (SGD) method. In the SR algorithm, the parameter update $\Delta\theta$ is obtained by solving the linear equation

$$S_{mn}\Delta\theta = -\eta g_\theta, \tag{10}$$

where $S_{mn}$ is approximated as

$$S_{mn} \approx \sum_{k \in \mathcal{V}} P(D_k) O^*_{k\theta_m} O_{k\theta_n} - \sum_{k \in \mathcal{V}} P(D_k) O^*_{k\theta_m} \sum_{k \in \mathcal{V}} P(D_k) O_{k\theta_n} + \lambda \delta_{mn}. \tag{11}$$

Here, $\eta$ denotes the step size and $\lambda$ is a regularization parameter that stabilizes the optimization process. Generally, larger values of $\lambda$ lead to parameter updates that align more with the steepest descent direction[37, 38].

## 3. ALGORITHMIC IMPROVEMENTS

**3.1. Deterministic Configuration Selection.** Despite the success of MC sampling in optimizing NQS, it faces several challenges when applied to chemical systems. In the ground-state wave functions of molecular systems, the probabilities of electron configurations can span many orders of magnitude, with peaks typically around the Hartree-Fock and neighboring excited states. This characteristic makes the standard uniform MC sampling inefficient in capturing chemically relevant configurations that are less likely to occur. Moreover, MC sampling introduces statistical instability to NQS optimization, which leads to fluctuations in energy or configuration coefficients across iterations and can be mitigated by using deterministically generated samples[17]. Therefore, an alternative method that efficiently and deterministically generates a configuration sample $\mathcal{V}$ containing most of the chemically important configurations would be beneficial.

Here we found that the iterative construction scheme of the SCI method can be adapted to construct such a configuration sample $\mathcal{V}$ during NQS optimization with efficiency. In typical SCI method[39], the full configuration space is divided into three parts: the core space $\mathcal{V}$, which includes selected significant configurations or determinants; the connected space $\mathcal{C}$, which contains all other configurations connected to the current core space through Hamiltonian matrix elements with nonzero or sufficiently large amplitudes; and the remaining part of the configuration space. The important configurations in connected space $\mathcal{C}$ are iteratively selected and added into core space $\mathcal{V}$ during the SCI optimization.

In the NQS optimization, we note that a similar pattern, including $\mathcal{V}$, $\mathcal{C}$ and the rest part, is inherently present, as in SCI. By considering the required configuration sample $\mathcal{V}$ in NQS as the core space $\mathcal{V}$ similar to the SCI method, the connected space $\mathcal{C}$ can be automatically generated during the calculation of NQS energy. Besides, the configuration sample $\mathcal{V}^n = \{|D_k\rangle\}$ in NQS at a given iteration $n$ can be deterministically updated with the selection of the significant configurations from its connected space $\mathcal{C}^n$ based on the modulus of the NQS ansatz $|\Psi_\theta\rangle$ as follows:

a. **Calculate the local energy of each configuration in the current sample $\mathcal{V}^n$ as required for NQS optimization, while simultaneously selecting important configurations from its connected space $\mathcal{C}^n$.** As illustrated in Figure 2, when computing the local energy $E_{loc}(D_k) = \sum_\mu \frac{\langle D_\mu | \Psi_\theta \rangle}{\langle D_k | \Psi_\theta \rangle} \langle D_k | \widehat{H} | D_\mu \rangle$ of each configuration $|D_k\rangle$ in sample $\mathcal{V}^n$, all configurations denoted by $|D_\mu\rangle$ are generated, and their coefficients $\langle D_\mu | \Psi_\theta \rangle$ are evaluated. These configurations precisely constitute the connected space $\mathcal{C}^n$ of sample $\mathcal{V}^n$. Therefore, the important configurations from the connected space $\mathcal{C}^n$ can be selected by ensuring that their coefficients are greater than the selection cutoff $\epsilon$ during the energy calculation. Here, the NQS ansatz is an intermediate normalized form where the largest $|\langle D_k | \Psi_\theta \rangle|$ in sample $\mathcal{V}^n$ is set to one, and any $|D_\mu\rangle$ already present in sample $\mathcal{V}^n$ will be discarded.

b. **Update the sample $\mathcal{V}^n$ to $\mathcal{V}^{n+1}$ by incorporating these deterministically selected important configurations in $\mathcal{C}^n$.** Combine the configurations in the current sample $\mathcal{V}^n$ that satisfy $|\langle D_k | \Psi_\theta \rangle| > \epsilon$ with the selected configurations from the connected space $\mathcal{C}^n$ to construct a new sample $\mathcal{V}^{n+1}$. Then, optimize the NQS ansatz $|\Psi_\theta\rangle$ with this new sample. Generally speaking, most of the configurations in $\mathcal{V}^{n+1}$ come from $\mathcal{V}^n$, indicating that their local energies have already been calculated in step a.

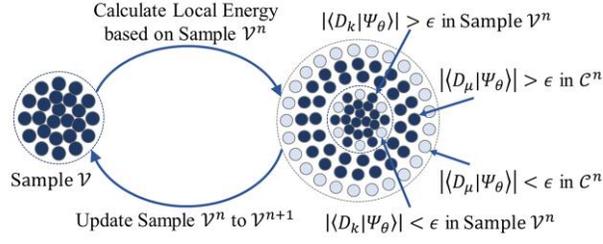

**Figure 2.** Graphical depiction of the deterministic selection process in configuration space. The dots represent the electron configurations. At a given iteration $n$, the current sample $\mathcal{V}$ (the small circle encompassing dark blue dots) is denoted as $\mathcal{V}^n$. During the calculation of local energies for all configurations in sample $\mathcal{V}^n$, its connected space $\mathcal{C}^n$ (the annular zone) is automatically generated. Configurations in both $\mathcal{V}^n$ and $\mathcal{C}^n$ are then divided into two sets based on whether their moduli $|\langle D_k|\Psi_\theta\rangle|$ and $|\langle D_\mu|\Psi_\theta\rangle|$ exceed the cutoff $\epsilon$. Those configurations whose moduli are larger than $\epsilon$ constitute the updated new sample $\mathcal{V}^{n+1}$.

During the iterative optimization of NQS, the initial configuration sample $\mathcal{V}$ consists of one MC sample and is deterministically updated at each iteration. Since the NQS ansatz typically undergoes slight and continuous changes after each optimization iteration, the important configurations $|D_k\rangle$ that should be included in the sample $\mathcal{V}$ also vary in a similar manner. Therefore, the aforementioned strategy ensures that the sample $\mathcal{V}$ maintains the inclusion of all desired configurations satisfying $|\langle D_k|\Psi_\theta\rangle| > \epsilon$ after the initial few iterations. Moreover, the size of the sample $\mathcal{V}$ can be conveniently adjusted by modifying the magnitude of $\epsilon$, where setting $\epsilon = 0$ corresponds to selecting the entire configuration space. It is worth noting that, for instance, an exact selection with $\epsilon = 10^{-6}$ is comparable to performing MC sampling of the square modulus with a total of $10^{12}$ configurations but without the presence of noise. Because in both cases, the aim is to generate all configurations with a probability exceeding $10^{-12}$.

**3.2. Non-stochastic Optimization of NQS.** Here, we describe how to incorporate the deterministic selection scheme in the configuration space into the optimization of NQS. As mentioned in Section 2.2, accurately evaluating the probability $P(D_k)$ of each configuration in eq 5 is necessary for computing the energy of NQS. Due to the intractable summation over full configuration space in the denominator of accurate $P(D_k)$, a statistical approximation $P(D_k) \approx \frac{N(D_k)}{N_{tot}}$ based on MC sampling is typically employed. In the case of using the deterministically selected sample, a new approximation that does not rely on MC sampling becomes necessary. Therefore, we proposed using $P(D_k) \approx \frac{|\langle D_k|\Psi_\theta\rangle|^2}{\sum_{i\in\mathcal{V}}|\langle D_i|\Psi_\theta\rangle|^2}$, where we only consider configurations that are included in the configuration sample $\mathcal{V}$, as other terms $|\langle D_i|\Psi_\theta\rangle|^2$ in the summation are small enough to be omitted. With this new approximation, the energy expectation value of NQS can be derived as

$$E_\theta \approx \sum_{k\in\mathcal{V}} \frac{|\langle D_k|\Psi_\theta\rangle|^2}{\sum_{i\in\mathcal{V}}|\langle D_i|\Psi_\theta\rangle|^2} E_{loc}(D_k). \tag{12}$$

Similarly, other expectation values can be estimated. While it may be difficult to determine whether the energy in eq 12 remains variational, we can always obtain a variational energy value using

$$E_\theta = \frac{\langle\Psi_\theta|\hat{H}|\Psi_\theta\rangle}{\langle\Psi_\theta|\Psi_\theta\rangle} \approx \frac{\sum_{ij\in\mathcal{V}}\langle\Psi_\theta|D_i\rangle\langle D_i|\hat{H}|D_j\rangle\langle D_j|\Psi_\theta\rangle}{\sum_{k\in\mathcal{V}}|\langle D_k|\Psi_\theta\rangle|^2}, \tag{13}$$

which approximates the wave function $|\Psi_\theta\rangle$ with the selected sample $\mathcal{V}$. As demonstrated in Section 4.2, the energies obtained from eqs 12 and 13 are usually equal with high precision. Thus, it is safe to use the energy from eq 12 directly.

To illustrate the changes in the optimization process after replacing the MC sample with the deterministically selected sample, we provide a schematic of the two optimization schemes in Figure 3. The most important difference between these two schemes is that our newly developed non-stochastic *selected configuration* (SC) iteration scheme avoids the time-consuming MC sampling in each iteration loop and combines the update of the configuration sample $\mathcal{V}$ with the energy evaluation. This improvement leads to a notable enhancement in the speed of NQS optimization. Furthermore, the non-stochastic SC iteration scheme provides a more stable optimization process compared to the stochastic MC iteration scheme.

The SC iteration scheme is expected to result in faster and more stable optimization of NQS, as demonstrated in Section 4.1. In practice, we can control the size of the selected sample by adjusting the magnitude of the cutoff $\epsilon$. For large chemical systems, a larger value of $\epsilon$ is recommended in the initial iterations, as during the optimization of a randomly initialized NQS ansatz in the early iterations, the NQS ansatz often deviates significantly from the ground-state wave function, resulting in very large selected configuration samples. Additionally, the MC iteration scheme with a limited sample size can be employed to pre-optimize the randomly initiated NQS ansatz, enabling the NQS ansatz to resemble the ground-state wave function and making it suitable for further optimization using the SC iteration scheme.

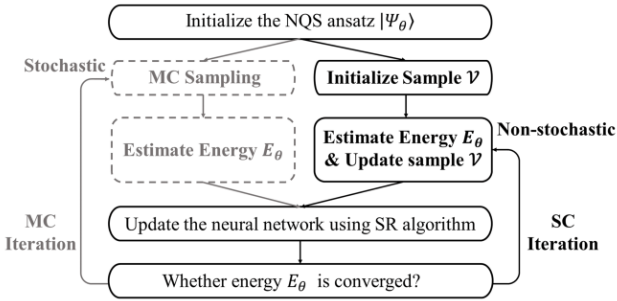

**Figure 3.** Schematic diagram of two types of optimization schemes for NQS. In the second and third steps, the two gray blocks on the left belong to the stochastic MC iteration scheme which involves MC sampling at each iteration, while the two black blocks on the right are part of the non-stochastic selected configuration (SC) iteration scheme which utilizes the selected configuration sample during energy estimation. Due to the incorporation of configuration selection into energy estimation in the SC iteration scheme, update the configuration sample costs minimal additional time. Therefore, the SC iteration scheme consists of only three steps per iteration, while the MC scheme requires four steps.

## 4. RESULTS

**4.1. Comparison with Stochastic Iteration Scheme.** We compared the performance of the non-stochastic SC iteration scheme with the stochastic iteration scheme by calculating the ground-state energies of various molecules in the minimal STO-3G basis set. The results in Table 1 demonstrate that the SC iteration scheme consistently achieves remarkable accuracy for all molecules considered, exhibiting its capability to effectively select important configurations in large configuration spaces. The results are compared with previously reported energies optimized through stochastic sampling. In Table 1, the first five molecules were calculated in ref 16 using the restricted Boltzmann machine based NQS through standard MC sampling, while the last fourteen molecules were studied in ref 22 by optimizing a large autoregressive NQS through stochastically sampling with an equivalent total configuration number of up to $N_{tot} = 10^{12}$. Therefore, in our SC iteration scheme, we employed the same hidden unit density $\alpha$ as in ref 16 and a selection cutoff of $\epsilon = 10^{-5}$ for the first five molecules. For the last fourteen molecules, we set $\alpha = 4$ and $\epsilon = 10^{-6}$. The results indicate that the SC scheme outperforms the standard MC sampling with the same wavefunction ansatz for these five molecules, and achieves comparable accuracy to the autoregressive sampling for the last fourteen molecules as both methods incorporate sufficient relevant configurations. Importantly, the SC scheme is applicable to any types of NQS, while the autoregressive sampling is limited to autoregressive neural networks.

The convergence differences between our non-stochastic SC iteration scheme and the standard MC iteration scheme are demonstrated in Figure 4. Both methods achieve high accuracy in calculating the ground-state energies of the N$_2$ molecule at its equilibrium bond length of 1.19 Å and the stretched bond length of 2.10 Å in the STO-3G basis set. The non-stochastic SC iteration scheme exhibits faster energy convergence and yields lower converged energy values compared to the stochastic MC iteration scheme. Additionally, the SC method consistently drops the energy and demonstrates smooth convergence curves, whereas the stochastic MC iteration scheme shows energy reduction with prominent oscillations. These improvements can be primarily attributed to the SC iteration scheme's ability to select more important configurations without introducing noise.

Figure S1 illustrates the percentages of generated configurations relative to the full configuration space with these two iteration schemes. Initially, both schemes generate configurations that cover the entire configuration space, and then the generated configuration sample gradually reduces to a stable and small proportion of the full configuration space. With a selection cutoff of $\epsilon = 10^{-6}$, the SC iteration scheme selects approximately 15% of the configurations, while the MC iteration scheme samples around 3% of the configurations with a total number of $N_{tot} = 10^6$, from the 14,400 configurations of the N$_2$ molecule.

The improved speed of the non-stochastic SC iteration scheme is another highlight of this approach, as illustrated in Figure 5, which can be attributed to the avoidance of time-consuming MC sampling. The practical computational cost of the SC scheme arises during energy calculation and scales with the number of selected unique configurations, while the MC scheme incurs additional cost for MC sampling, which scales with the overall sample size.

**Table 1.** The molecular ground-state energies (in Ha) obtained using various methods. The "Stochastic Sampling" column correspond to the reported energies of NQS achieved through stochastic sampling of configurations, while the "Non-stochastic Selection" column shows the results obtained using the SC iteration scheme. Additionally, the table includes the total number of electron configurations and the selected percentages at the end of the SC iteration scheme.

| Molecule | Total Configurations | Selected Percentage | CCSD(T) | Stochastic Sampling[a] | Non-stochastic Selection | FCI |
|---|---|---|---|---|---|---|
| LiH | 225 | 63.11% | −7.8828 | −7.8826 | **−7.8828** | −7.8828 |
| $H_2O$ | 441 | 46.03% | −75.0232 | −75.0232 | **−75.0233** | −75.0233 |
| $NH_3$ | 3,136 | 18.65% | −55.5281 | −55.5277 | **−55.5279** | −55.5282 |
| $N_2$ | 14,400 | 6.81% | −107.6738 | −107.6767 | **−107.6772** | −107.6774 |
| $C_2$ | 44,100 | 1.95% | −74.6876 | −74.6892 | **−74.6902** | −74.6908 |
| LiH | 225 | 70.22% | −7.7845 | **−7.7845** | **−7.7845** | −7.7845 |
| $H_2O$ | 441 | 63.72% | −75.0155 | **−75.0155** | −75.0155 | −75.0155 |
| $CH_2$ | 735 | 51.70% | −37.5044 | **−37.5044** | −37.5044 | −37.5044 |
| $O_2$ | 1,200 | 48.17% | −147.7485 | −147.7500 | **−147.7502** | −147.7502 |
| $BeH_2$ | 1,225 | 44.73% | −14.4729 | **−14.4729** | −14.4729 | −14.4729 |
| $H_2S$ | 3,025 | 27.17% | −394.3546 | **−394.3546** | −394.3546 | −394.3546 |
| $NH_3$ | 3,136 | 46.01% | −55.5210 | **−55.5211** | −55.5211 | −55.5211 |
| $N_2$ | 14,400 | 17.44% | −107.6579 | −107.6595 | **−107.6601** | −107.6602 |
| $CH_4$ | 15,876 | 27.27% | −39.8062 | **−39.8062** | −39.8060 | −39.8063 |
| $C_2$ | 44,100 | 7.98% | −74.6876 | −74.6899 | **−74.6904** | −74.6908 |
| LiF | 44,100 | 7.25% | −105.1663 | **−105.1662** | −105.1661 | −105.1662 |
| $PH_3$ | 48,400 | 8.29% | −338.6984 | **−338.6984** | −338.6983 | −338.6984 |
| LiCl | 1,002,001 | 0.55% | −460.8500 | **−460.8496** | −460.8496 | −460.8496 |
| $Li_2O$ | 41,409,225 | 0.04% | −87.8931 | −87.8909 | **−87.8918** | −87.8927 |

[a]The first five molecules were calculated in ref 16 using the restricted Boltzmann machine based NQS, while the last fourteen molecules were studied in ref 22 by optimizing a large autoregressive NQS.

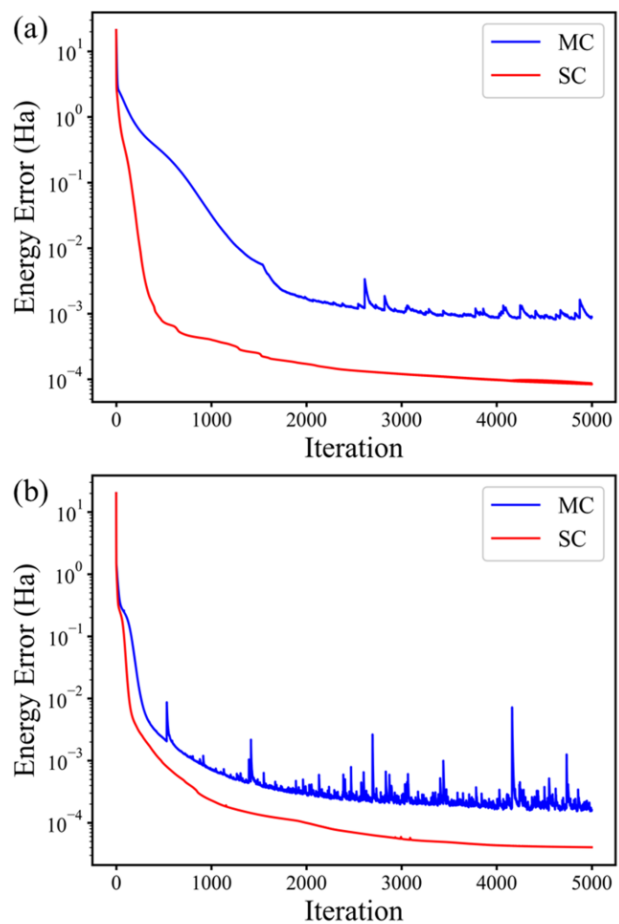

**Figure 4.** The convergence performances of the standard MC iteration scheme and the non-stochastic SC iteration scheme on the $N_2$ molecule. The upper plot shows the convergence at the equilibrium bond length of 1.19 Å, while the lower plot represents the convergence at the stretched bond length of 2.10 Å. Both schemes achieve accurate ground-state energy calculations for the $N_2$ molecule in the STO-3G basis set. The energy error is relative to the FCI results.

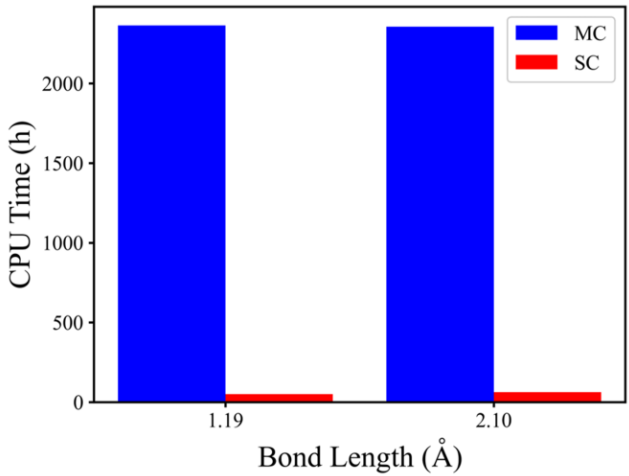

**Figure 5.** The CPU time in hours required by standard MC iteration scheme and non-stochastic SC iteration scheme for the $N_2$ molecule at its equilibrium bond length of 1.19 Å and the stretched bond length of 2.10 Å in STO-3G basis set.

**4.2. Potential Energy Curves of Carbon Dimer with Different Selection Cutoffs $\epsilon$.** In this section, we investigate the performance of the non-stochastic SC iteration scheme by varying the selection cutoff $\epsilon$ on the potential energy curves of the ground-state carbon dimer $C_2$ in the minimal STO-3G basis set. The carbon dimer $C_2$ is a diatomic molecule with strong correlations and multireference characteristics. We employ six selection cutoffs ranging from $10^{-8}$ to $10^{-3}$ in the SC iteration scheme, and all of them yield potential energy curves that closely match the FCI curve across the entire dissociation range as shown in Figure 6a. In contrast, the CCSD(T) method fails when the bond length is elongated. Moreover, except for the largest cutoff of $\epsilon = 10^{-3}$, the NQS ansatz optimized through the SC iteration scheme outperforms the CCSD(T) method near the equilibrium bond lengths, where CCSD(T) performs well. Figure 6b demonstrates that all cutoffs, except $\epsilon = 10^{-3}$, achieve comparable high accuracy at all considered bond lengths. The assumption that lower selection cutoffs result in higher energy accuracy is not valid since $\epsilon = 10^{-4}$, even though a larger configuration sample $\mathcal{V}$ is generated when using lower cutoffs, as depicted in Figures 6c and S2. Therefore, the energy accuracy with small selection cutoffs is primarily limited by the stochastic reconfiguration parameter optimization algorithm and the expressive power of the restricted Boltzmann machine based NQS ansatz with a hidden unit density $\alpha$ of 2.

The percentage of the selected configuration sample $\mathcal{V}$ relative to the full configuration space of the $C_2$ molecule in STO-3G basis set not only increases with decreasing selection cutoff, but also increases more rapidly as the cutoff becomes smaller. The disparity in selected percentages between bond lengths also becomes more pronounced as the selection cutoff decreases. Although there are limitations on energy accuracy, larger configuration samples with lower selection cutoffs can provide more robust convergence, as depicted in Figure 6b. The hollow dots indicate that these optimizations are not converged and that the results represent the best energies during the optimization process, rather than the converged energies. Here, convergence is achieved when the NQS energy changes by less than $10^{-6}$ Ha for ten consecutive iterations.

We also explore the differences between the energies obtained during the optimization of NQS using eq 12 and the variational energies of the optimized NQS ansatzes using eq 13. The results in Figure 6d suggest that NQS ansatzes optimized with lower selection cutoff $\epsilon$ exhibit smaller energy deviations. When a selection cutoff equal to or less than $10^{-5}$ is used, the energy deviations reduce to less than $10^{-6}$ Ha, suggesting that the energy obtained through eq 12 can be considered as variational.

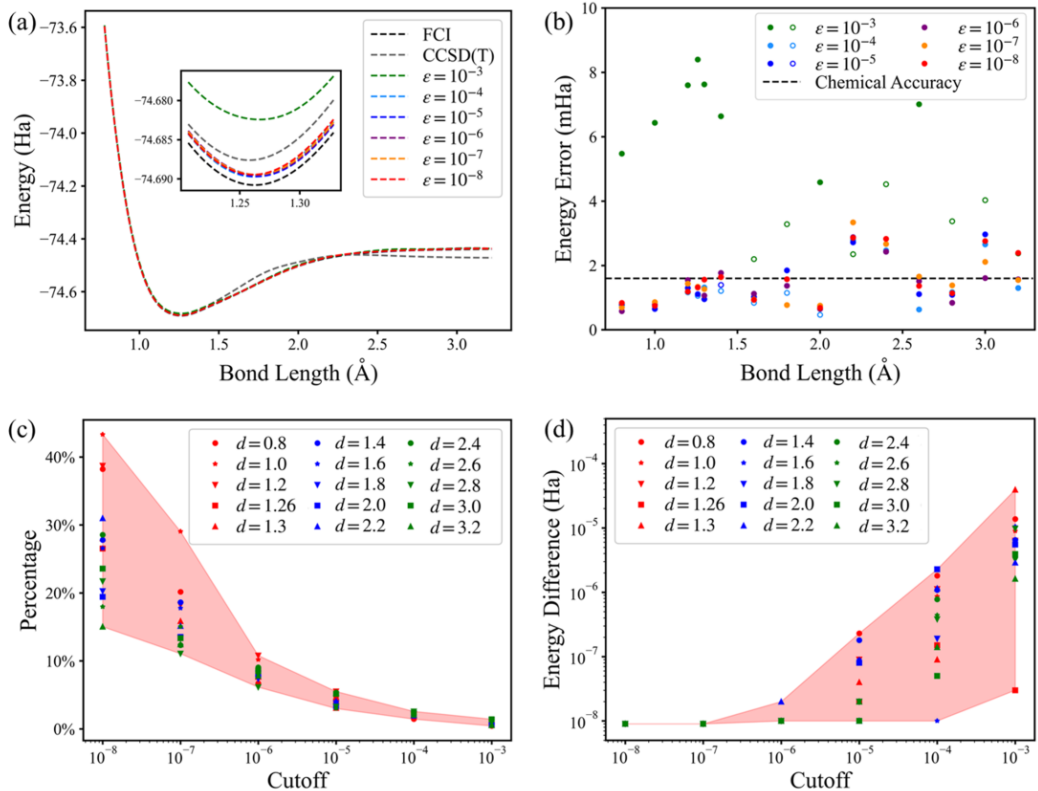

**Figure 6.** The performances of the SC iteration scheme on calculating the ground-state $C_2$ molecule in STO-3G basis set. (a) Potential energy curves obtained through the CCSD(T) method, the FCI method and the NQS optimized using the SC iteration scheme with six selection cutoffs $\epsilon$. (b) Differences between the FCI results and the energies obtained using different selection cutoff $\epsilon$ at various bond lengths. The hollow dots indicate that these optimizations are not converged and that the results represent the best energies during the optimization process, rather than the converged energies. (c) Percentage of the selected configuration sample relative to the full configuration space at various bond lengths $d$ using different selection cutoffs $\epsilon$. (d) Differences between the energies obtained during the optimization of NQS using eq 12 and the variational energies of the optimized NQS ansatzes using eq 13 with different selection cutoffs $\epsilon$ at various bond lengths $d$.

**4.3. Performance in the Equidistant Linear Hydrogen Chain Systems.** In the minimal STO-6G basis set, the equidistant linear hydrogen chain resembles a one-dimensional Hubbard model as there is only one orbital per hydrogen atom and exhibits strong electron correlations[40]. This molecule serves as a convenient system to investigate the size-dependence of the non-stochastic SC iteration scheme for NQS and has been recently utilized to evaluate various quantum chemical methods[40]. Moreover, from a quantum chemical perspective[41, 42], the equidistant linear hydrogen chain is an elongated molecule for which the localized molecular orbital (LMO) basis, such as the one obtained with the Boys-localization[43] in PySCF packages[44, 45], provides a more accurate description of the ground state than the canonical molecular orbital (CMO) basis. Hence, this system allows for the examination of the performance disparity between NQS with LMOs and CMOs, as the NQS ansatz is not invariant under orbital rotations[17, 46].

We investigate the performance of the non-stochastic SC iteration scheme for NQS on equidistant linear hydrogen chains containing 4, 6, 8, 10, and 12 hydrogen atoms respectively. The ground states of these chains were studied at both the near equilibrate H-H bond length of 1.8 Bohr and the elongated H-H bond length of 3.6 Bohr. As depicted in Figures 7a and 7b, NQS with CMOs exhibits a noticeable increase in energy errors with system size at both bond lengths, while NQS with LMOs consistently achieves remarkable accuracy. Figure 7a also demonstrates that, when using LMOs, the NQS optimized with the SC iteration scheme outperforms the CCSD(T) method at the near equilibrate H-H bond length of 1.8 Bohr. The result of CCSD(T) is not presented in Figure 7b, due to its failure in these elongated systems as shown in Table S2.

To strike a balance between the accuracy of converged NQS energy and the size of selected configuration sample, we set the selection cutoff $\epsilon$ to be $10^{-5}$ for CMOs and $10^{-3}$ for LMOs. Figures 7c and 7d illustrate the percentage of the selected sample relative to the full configuration space for these optimized NQS ansatz and provide a comparison with their corresponding FCI wave functions. The number of configurations in the selected sample is presented in Figure S3. Notably, when using LMOs, the excellent match between the sample size of the NQS ansatz and the FCI wave function aligns with the remarkable energy accuracy achieved, thereby suggesting the success of the SC iteration scheme in selecting all crucial configurations despite the increasing system size. Furthermore, for the equidistant linear hydrogen chain $H_{12}$ containing 12 atoms at an H-H bond length of 1.8 Bohr, the RBM-based NQS ansatz demonstrates its remarkable expressive power by fitting the probability distribution of over 400,000 selected configurations with LMOs. Conversely, when using CMOs, a disparity in the size of the selected configuration samples between the NQS ansatz and the FCI wave function is observed, highlighting the challenge faced by the RBM-based NQS in approximating the ground-state wave function of the equidistant linear hydrogen chain in the CMO basis using the stochastic reconfiguration parameter optimization algorithm.

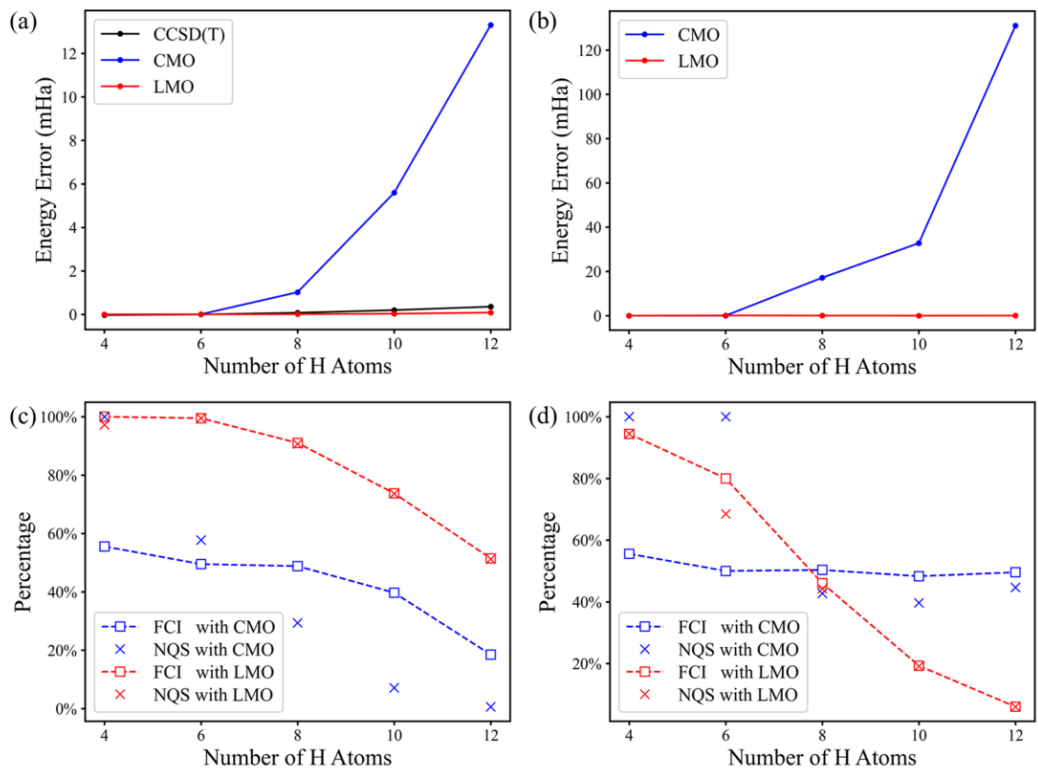

**Figure 7.** The performances of the NQS optimized through the SC iteration scheme in calculating the ground-states of equidistant linear hydrogen chain systems in the STO-6G basis set using both the LMO and CMO basis, exploring two H-H bond lengths of 1.8 Bohr (left) and 3.6 Bohr (right). (a) Energy errors relative to the FCI results at the H-H bond length of 1.8 Bohr. (b) Energy errors relative to the FCI results at the H-H bond length of 3.6 Bohr. The result of CCSD(T) is not presented due to its failure in these elongated systems. (c) Percentages of the selected sample relative to the full configuration space using the FCI wave function and the optimized NQS ansatz at the H-H bond length of 1.8 Bohr. (d) Percentages of the selected sample relative to the full configuration space using the FCI wave function and the optimized NQS ansatz at the H-H bond length of 3.6 Bohr.

The rapidly increasing size of the selected configuration sample shown in Figure S3 reveals a limitation of the NQS methods. When optimizing the NQS ansatz, it is crucial to have a relatively small configuration sample generated from either the MC approach or the SC approach in order to efficiently approximate the NQS energy and its gradient with respect to neural-network parameters. However, the required size of the configuration sample depends on the scarcity of configurations that contribute significantly to the FCI wave function. For instance, as shown in Table 1, the Li$_2$O molecule with 41,409,225 entire configurations only requires 15,001 relevant configurations (about 0.04% relative to the full space) to accurately approximate the exact ground state. In contrast, the H$_{12}$ molecule with 853,776 entire configurations at an H-H bond length of 1.8 Bohr requires 433,473 configurations (about 50.77% relative to the full space). Although this limitation is system-dependent, it can be mitigated by trading off accuracy and optimization convergence. For example, increasing the selection cutoff to $10^{-2}$ for the H$_{12}$ molecule at an H-H bond length of 1.8 Bohr in LMOs reduces the size of the selected configuration sample to 171,751 (about 20.12% of the full space), while the energy of NQS increases by 0.43 mHa and become slightly higher than the CCSD(T) result. The final energy oscillation during the optimization of NQS ansatz is less than $5 \times 10^{-6}$ Ha.

## 5. CONCLUSION

In this study, we present a non-stochastic optimization algorithm for neural-network quantum states, which is based on a generalized exact selection scheme employing the modulus of the NQS. This algorithm allows for the deterministic update of the important configuration sample in parallel with the energy evaluation, leading to minimal additional computational cost. When compared with two other stochastic optimization methods, we find that this non-stochastic optimization algorithm results in similar or even better accuracy in calculating the ground-state energies of the nineteen molecules previously examined. Additionally, this non-stochastic optimization scheme yields faster and smoother convergence than standard Monte Carlo methods.

Furthermore, we have investigated the behavior of NQS under different orbital rotations, specifically comparing the use of CMOs and LMOs in equidistant linear hydrogen chain systems. The results indicate that the NQS utilizing CMOs showcases a significant increase in energy errors with system sizes, while NQS employing LMOs consistently achieves remarkable accuracy. Therefore, it is advisable and feasible to integrate the molecular orbital optimization into the optimization of NQS[47], enhancing its numerical flexibility and making it more efficient in finding the optimal ground-state energy in various systems. Furthermore, the exploration of other powerful neural-network architectures[28, 48-51] or parameter optimization algorithms[52-54] warrants further attention, which will shed light on the applications of NQS in chemical systems.

## ASSOCIATED CONTENT

### Supporting Information

Computational details of the test calculations, plots of the percentages of the selected sample relative to the full configuration space for both the $N_2$ and $C_2$ molecule, and the figure of the size of the selected configuration samples in equidistant linear hydrogen chain systems (PDF)

## AUTHOR INFORMATION

**Corresponding Author**

*E-mail: hshu@mail.tsinghua.edu.cn.

## ACKNOWLEDGMENT

This work was financially supported by the National Natural Science Foundation of China (Grant Nos. 22222605 and 22076095) and the Tsinghua Xuetang Talents Program. Mr. Xiang Li thanks Dr. Junjie Song of Tsinghua University, Prof. Zhendong Li of Beijing Normal University for helpful discussions.

## REFERENCES

(1) Huron, B.; Malrieu, J. P.; Rancurel, P., Iterative perturbation calculations of ground and excited state energies from multiconfigurational zeroth-order wavefunctions. *J. Chem. Phys.* **1973,** *58* (12), 5745-5759.

(2) Greer, J. C., Estimating full configuration interaction limits from a Monte Carlo selection of the expansion space. *J. Chem. Phys.* **1995,** *103* (5), 1821-1828.

(3) Holmes, A. A.; Tubman, N. M.; Umrigar, C. J., Heat-Bath Configuration Interaction: An Efficient Selected Configuration Interaction Algorithm Inspired by Heat-Bath Sampling. *J. Chem. Theory Comput.* **2016,** *12* (8), 3674-3680.

(4) Liu, W.; Hoffmann, M. R., iCI: Iterative CI toward full CI. *J. Chem. Theory Comput.* **2016,** *12* (3), 1169-1178.

(5) Schriber, J. B.; Evangelista, F. A., Communication: An adaptive configuration interaction approach for strongly correlated electrons with tunable accuracy. *J. Chem. Phys.* **2016,** *144* (16), 161106.

(6) Tubman, N. M.; Lee, J.; Takeshita, T. Y.; Head-Gordon, M.; Whaley, K. B., A deterministic alternative to the full configuration interaction quantum Monte Carlo method. *J. Chem. Phys.* **2016,** *145* (4), 044112.

(7) Sharma, S.; Holmes, A. A.; Jeanmairet, G.; Alavi, A.; Umrigar, C. J., Semistochastic Heat-Bath Configuration Interaction Method: Selected Configuration Interaction with Semistochastic Perturbation Theory. *J. Chem. Theory Comput.* **2017,** *13* (4), 1595-1604.

(8) Coe, J. P., Machine Learning Configuration Interaction. *J. Chem. Theory Comput.* **2018,** *14* (11), 5739-5749.

(9) Coe, J. P., Machine Learning Configuration Interaction for ab Initio Potential Energy Curves. *J. Chem. Theory Comput.* **2019,** *15* (11), 6179-6189.

(10) Tubman, N. M.; Freeman, C. D.; Levine, D. S.; Hait, D.; Head-Gordon, M.; Whaley, K. B., Modern Approaches to Exact Diagonalization and Selected Configuration Interaction with the Adaptive Sampling CI Method. *J. Chem. Theory Comput.* **2020,** *16* (4), 2139-2159.

(11) Zhang, N.; Liu, W.; Hoffmann, M. R., Iterative Configuration Interaction with Selection. *J. Chem. Theory Comput.* **2020,** *16* (4), 2296-2316.

(12) Goings, J. J.; Hu, H.; Yang, C.; Li, X., Reinforcement Learning Configuration Interaction. *J. Chem. Theory Comput.* **2021,** *17* (9), 5482-5491.

(13) Herzog, B.; Casier, B.; Lebègue, S.; Rocca, D., Solving the Schrödinger Equation in the Configuration Space with Generative Machine Learning. *J. Chem. Theory Comput.* **2023,** *19* (9), 2484-2490.

(14) Freericks, J. K.; Nikolić, B. K.; Frieder, O., The nonequilibrium quantum many-body problem as a paradigm for extreme data science. *Int. J. Mod. Phys. B* **2014,** *28* (31), 1430021.

(15) Carleo, G.; Troyer, M., Solving the quantum many-body problem with artificial neural networks. *Science* **2017,** *355* (6325), 602-606.

(16)	Choo, K.; Mezzacapo, A.; Carleo, G., Fermionic neural-network states for ab-initio electronic structure. *Nat. Commun.* **2020,** *11* (1), 2368.

(17)	Yang, P.-J.; Sugiyama, M.; Tsuda, K.; Yanai, T., Artificial Neural Networks Applied as Molecular Wave Function Solvers. *J. Chem. Theory Comput.* **2020,** *16* (6), 3513-3529.

(18)	Hermann, J.; Schätzle, Z.; Noé, F., Deep-neural-network solution of the electronic Schrödinger equation. *Nat. Chem.* **2020,** *12* (10), 891-897.

(19)	Pfau, D.; Spencer, J. S.; Matthews, A. G. D. G.; Foulkes, W. M. C., Ab initio solution of the many-electron Schrodinger equation with deep neural networks. *Phys. Rev. Res.* **2020,** *2* (3), 033429.

(20)	Hermann, J.; Spencer, J.; Choo, K.; Mezzacapo, A.; Foulkes, W. M. C.; Pfau, D.; Carleo, G.; Noé, F., Ab-initio quantum chemistry with neural-network wavefunctions. *arXiv* **2022,** *2208.12590*.

(21)	Sharir, O.; Levine, Y.; Wies, N.; Carleo, G.; Shashua, A., Deep Autoregressive Models for the Efficient Variational Simulation of Many-Body Quantum Systems. *Phys. Rev. Lett.* **2020,** *124* (2), 020503.

(22)	Barrett, T. D.; Malyshev, A.; Lvovsky, A. I., Autoregressive neural-network wavefunctions for ab initio quantum chemistry. *Nat. Mach. Intell.* **2022,** *4* (4), 351-358.

(23)	Zhao, T.; Stokes, J.; Veerapaneni, S., Scalable neural quantum states architecture for quantum chemistry. *Mach. Learn.: Sci. Technol.* **2023,** *4* (2), 025034.

(24)	Foulkes, W. M. C.; Mitas, L.; Needs, R. J.; Rajagopal, G., Quantum Monte Carlo simulations of solids. *Rev. Mod. Phys.* **2001,** *73* (1), 33-83.

(25)	Austin, B. M.; Zubarev, D. Y.; Lester, W. A., Quantum Monte Carlo and Related Approaches. *Chem. Rev.* **2012,** *112* (1), 263-288.

(26)	Deng, D.-L.; Li, X.; Das Sarma, S., Machine learning topological states. *Phys. Rev. B* **2017,** *96* (19), 195145.

(27) Deng, D.-L.; Li, X.; Das Sarma, S., Quantum Entanglement in Neural Network States. *Phys. Rev. X* **2017,** *7* (2), 021021.

(28) Nomura, Y.; Darmawan, A. S.; Yamaji, Y.; Imada, M., Restricted Boltzmann machine learning for solving strongly correlated quantum systems. *Phys. Rev. B* **2017,** *96* (20), 205152.

(29) Choo, K.; Carleo, G.; Regnault, N.; Neupert, T., Symmetries and Many-Body Excitations with Neural-Network Quantum States. *Phys. Rev. Lett.* **2018,** *121* (16), 167204.

(30) Glasser, I.; Pancotti, N.; August, M.; Rodriguez, I. D.; Cirac, J. I., Neural-Network Quantum States, String-Bond States, and Chiral Topological States. *Phys. Rev. X* **2018,** *8* (1), 011006.

(31) Vieijra, T.; Casert, C.; Nys, J.; De Neve, W.; Haegeman, J.; Ryckebusch, J.; Verstraete, F., Restricted Boltzmann Machines for Quantum States with Non-Abelian or Anyonic Symmetries. *Phys. Rev. Lett.* **2020,** *124* (9), 097201.

(32) Chen, J.; Cheng, S.; Xie, H.; Wang, L.; Xiang, T., Equivalence of restricted Boltzmann machines and tensor network states. *Phys. Rev. B* **2018,** *97* (8), 085104.

(33) Li, S.; Pan, F.; Zhou, P.; Zhang, P., Boltzmann machines as two-dimensional tensor networks. *Phys. Rev. B* **2021,** *104* (7), 075154.

(34) Hastings, W. K., Monte Carlo Sampling Methods Using Markov Chains and Their Applications. *Biometrika* **1970,** *57* (1), 97-109.

(35) Zhang, H.; Mandic, D. P., Is a Complex-Valued Stepsize Advantageous in Complex-Valued Gradient Learning Algorithms? *IEEE Trans. Neural Netw. Learn. Syst.* **2016,** *27* (12), 2730-2735.

(36) Sorella, S.; Casula, M.; Rocca, D., Weak binding between two aromatic rings: Feeling the van der Waals attraction by quantum Monte Carlo methods. *J. Chem. Phys.* **2007,** *127* (1), 014105.

(37) Toulouse, J.; Umrigar, C. J., Optimization of quantum Monte Carlo wave functions by energy minimization. *J. Chem. Phys.* **2007,** *126* (8), 084102.

(38) Umrigar, C. J.; Toulouse, J.; Filippi, C.; Sorella, S.; Hennig, R. G., Alleviation of the Fermion-Sign Problem by Optimization of Many-Body Wave Functions. *Phys. Rev. Lett.* **2007,** *98* (11), 110201.

(39) Smith, J. E. T.; Mussard, B.; Holmes, A. A.; Sharma, S., Cheap and Near Exact CASSCF with Large Active Spaces. *J. Chem. Theory Comput.* **2017,** *13* (11), 5468-5478.

(40) Simons Collaboration on the Many-Electron, P.; Motta, M.; Ceperley, D. M.; Chan, G. K.-L.; Gomez, J. A.; Gull, E.; Guo, S.; Jiménez-Hoyos, C. A.; Lan, T. N.; Li, J.; Ma, F.; Millis, A. J.; Prokof'ev, N. V.; Ray, U.; Scuseria, G. E.; Sorella, S.; Stoudenmire, E. M.; Sun, Q.; Tupitsyn, I. S.; White, S. R.; Zgid, D.; Zhang, S., Towards the Solution of the Many-Electron Problem in Real Materials: Equation of State of the Hydrogen Chain with State-of-the-Art Many-Body Methods. *Phys. Rev. X* **2017,** *7* (3), 031059.

(41) Wouters, S.; Van Neck, D., The density matrix renormalization group for ab initio quantum chemistry. *Eur. Phys. J. D* **2014,** *68* (9), 272.

(42) Guther, K.; Anderson, R. J.; Blunt, N. S.; Bogdanov, N. A.; Cleland, D.; Dattani, N.; Dobrautz, W.; Ghanem, K.; Jeszenszki, P.; Liebermann, N.; Manni, G. L.; Lozovoi, A. Y.; Luo, H.; Ma, D.; Merz, F.; Overy, C.; Rampp, M.; Samanta, P. K.; Schwarz, L. R.; Shepherd, J. J.; Smart, S. D.; Vitale, E.; Weser, O.; Booth, G. H.; Alavi, A., NECI: N-Electron Configuration Interaction with an emphasis on state-of-the-art stochastic methods. *J. Chem. Phys.* **2020,** *153* (3), 034107.

(43) Foster, J. M.; Boys, S. F., Canonical Configurational Interaction Procedure. *Rev. Mod. Phys.* **1960,** *32* (2), 300-302.

(44) Sun, Q.; Berkelbach, T. C.; Blunt, N. S.; Booth, G. H.; Guo, S.; Li, Z.; Liu, J.; McClain, J. D.; Sayfutyarova, E. R.; Sharma, S.; Wouters, S.; Chan, G. K.-L., PySCF: the Python-based simulations of chemistry framework. *WIREs Comput. Mol. Sci.* **2018,** *8* (1), e1340.

(45) Sun, Q.; Zhang, X.; Banerjee, S.; Bao, P.; Barbry, M.; Blunt, N. S.; Bogdanov, N. A.; Booth, G. H.; Chen, J.; Cui, Z.-H.; Eriksen, J. J.; Gao, Y.; Guo, S.; Hermann, J.; Hermes, M. R.; Koh, K.; Koval, P.; Lehtola, S.; Li, Z.; Liu, J.; Mardirossian, N.; McClain, J. D.; Motta, M.; Mussard, B.; Pham, H. Q.; Pulkin, A.; Purwanto, W.; Robinson, P. J.; Ronca,

E.; Sayfutyarova, E. R.; Scheurer, M.; Schurkus, H. F.; Smith, J. E. T.; Sun, C.; Sun, S.-N.; Upadhyay, S.; Wagner, L. K.; Wang, X.; White, A.; Whitfield, J. D.; Williamson, M. J.; Wouters, S.; Yang, J.; Yu, J. M.; Zhu, T.; Berkelbach, T. C.; Sharma, S.; Sokolov, A. Y.; Chan, G. K.-L., Recent developments in the PySCF program package. *J. Chem. Phys.* **2020,** *153* (2), 024109.

(46) Hagai, M.; Sugiyama, M.; Tsuda, K.; Yanai, T., Artificial neural network encoding of molecular wavefunctions for quantum computing. *Digital Discovery* **2023,** *2* (3), 634-650.

(47) Moreno, J. R.; Cohn, J.; Sels, D.; Motta, M., Enhancing the Expressivity of Variational Neural, and Hardware-Efficient Quantum States Through Orbital Rotations. *arXiv* **2023,** *2302.11588*.

(48) Gao, X.; Duan, L.-M., Efficient representation of quantum many-body states with deep neural networks. *Nat. Commun.* **2017,** *8* (1), 662.

(49) Carleo, G.; Nomura, Y.; Imada, M., Constructing exact representations of quantum many-body systems with deep neural networks. *Nat. Commun.* **2018,** *9* (1), 5322.

(50) Bennewitz, E. R.; Hopfmueller, F.; Kulchytskyy, B.; Carrasquilla, J.; Ronagh, P., Neural Error Mitigation of Near-Term Quantum Simulations. *Nat. Mach. Intell.* **2022,** *4* (7), 618-624.

(51) Robledo Moreno, J.; Carleo, G.; Georges, A.; Stokes, J., Fermionic wave functions from neural-network constrained hidden states. *Proc. Natl. Acad. Sci.* **2022,** *119* (32), e2122059119.

(52) Martens, J.; Grosse, R., Optimizing neural networks with Kronecker-factored approximate curvature. *International Conference on Machine Learning;* JMLR, **2015**; pp 2408–2417.

(53) Sabzevari, I.; Sharma, S., Improved Speed and Scaling in Orbital Space Variational Monte Carlo. *J. Chem. Theory Comput.* **2018,** *14* (12), 6276-6286.

(54) Frank, J. T.; Kastoryano, M. J., Learning Neural Network Quantum States with the Linear Method. *arXiv* **2021,** *2104.11011*.

**Table of Contents**

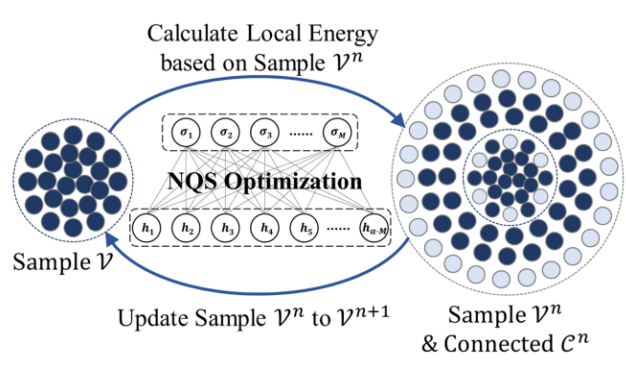

# A Non-stochastic Optimization Algorithm for Neural-network Quantum States

Xiang Li[1], Jia-Cheng Huang[1], Guang-Ze Zhang[1], Hao-En Li[1], Chang-Su Cao[1,2],

Dingshun Lv[2], Han-Shi Hu[1,*]

[1]*Department of Chemistry and Engineering Research Center of Advanced Rare-Earth Materials of Ministry of Education, Tsinghua University, Beijing 100084, China*

[2]*ByteDance Research, Zhonghang Plaza, No. 43, North 3rd Ring West Road, Haidian District, Beijing, 100089, China*

*Corresponding author. Email: hshu@mail.tsinghua.edu.cn

## 1. Computational Details

In this study, various standard quantum chemical methods such as Hartree-Fock (HF), coupled cluster with singles, doubles, and perturbatively corrected triples (CCSD(T)), and full configuration interaction (FCI) were utilized through the PySCF packages[1, 2]. Prior to optimizing the neural-network quantum states[3-5] (NQS), HF calculations were performed to obtain the canonical molecular orbital (CMO) and the necessary integrals. For cases where the localized molecular orbital (LMO) basis was used, Boys-localization[6] within the PySCF packages was employed.

## 2. Comparison with Stochastic Iteration Scheme.

The geometries for the first five molecules in Table 1 are sourced from the CCCBDB database[7], and the remaining fourteen from the PubChem database[8].

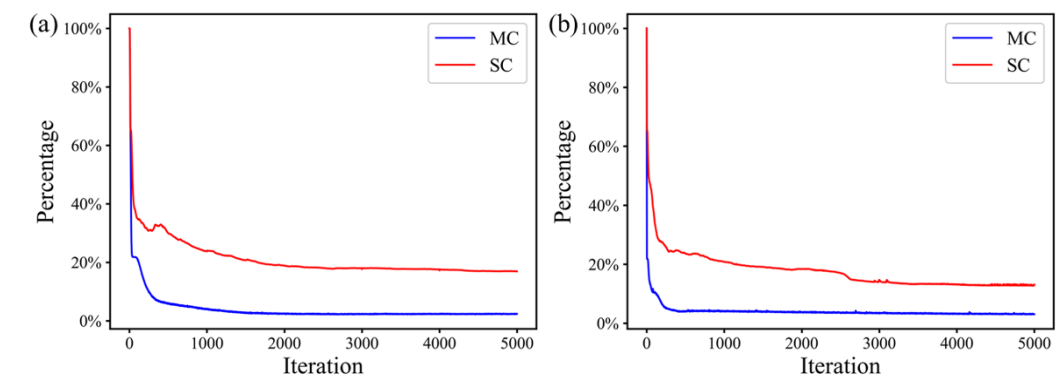

**Figure S1.** The percentages of the selected configuration sample relative to the full configuration space of the $N_2$ molecule at its equilibrium bond length of 1.19 Å (left) and the stretched bond length of 2.10 Å (right) in STO-3G basis set, using the standard MC iteration scheme and non-stochastic SC iteration scheme.

## 3. Carbon Dimer

The performance of the non-stochastic SC iteration scheme was evaluated by altering the selection cutoff $\epsilon$ on the potential energy curves of the ground-state carbon dimer in the minimal STO-3G basis set, with $\alpha$ and $\lambda$ set at 2 and $10^{-4}$ respectively. The percentiles of the selected configuration sample in relation to the full configuration space are illustrated in Figure S2.

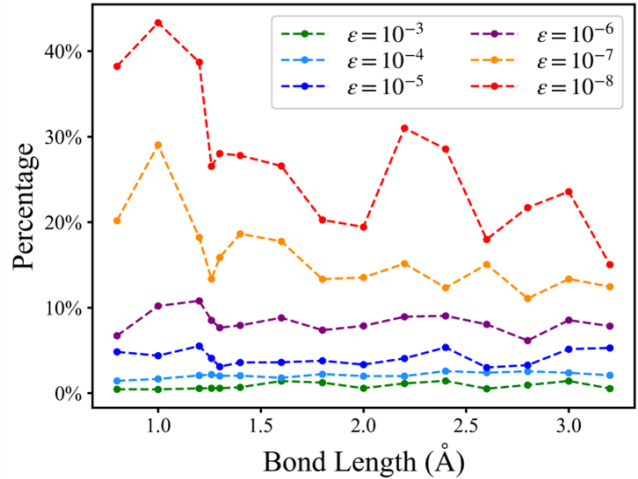

**Figure S2.** The percentages of the selected configuration sample relative to the full configuration space of the $C_2$ molecule in STO-3G basis set at various bond lengths $d$ using different selection cutoffs $\epsilon$.

## 4. Equidistant Linear Hydrogen Chain

In examining the performance of the non-stochastic SC iteration scheme in computing the ground-states of equidistant linear hydrogen chain systems in the STO-6G basis set using both the LMO and CMO basis, we utilized $\alpha = 4$, $\lambda = 10^{-5}$, $\epsilon = 10^{-5}$ for CMOs and $\epsilon = 10^{-3}$ for LMOs. Figure S3 presents the size of the selected sample, and Table S2 offers the ground-state energies obtained through various methods.

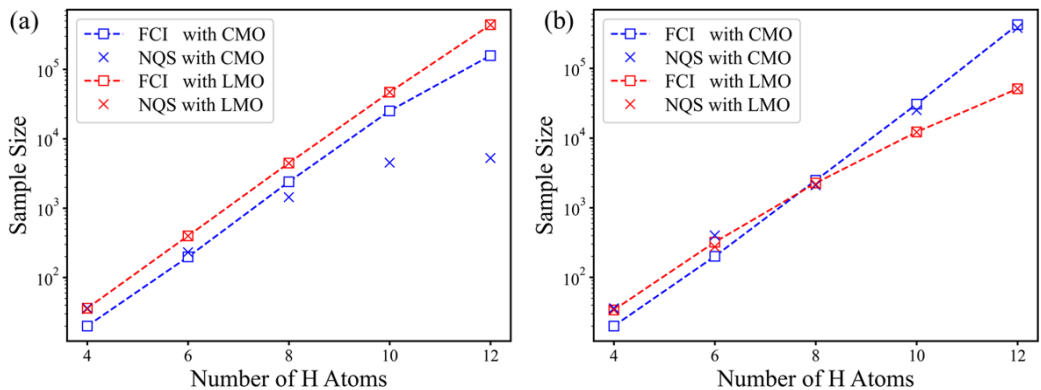

**Figure S3.** The selected sample size of the optimized NQS ansatz and the FCI wave function when calculating the ground-states of equidistant linear hydrogen chain systems in the STO-6G basis set using both the LMO and CMO basis, exploring two H-H bond lengths of 1.8 Bohr (left) and 3.6 Bohr (right). (a) The selected sample size at the H-H bond length of 1.8 Bohr. (b) The selected sample size at the H-H bond length of 3.6 Bohr.

Table S2. The ground-state energies obtained through the CCSD(T) method, the FCI method and the NQS optimized using the SC iteration scheme with both the LMO and CMO basis for equidistant linear hydrogen chains in the STO-6G basis at the H-H bond lengths of 1.8 and 3.6 Bohr.

| Molecule | CMO | LMO | CCSD(T) | FCI | CMO | LMO | CCSD(T) | FCI |
|---|---|---|---|---|---|---|---|---|
| | 1.8 Bohr | | | | 3.6 Bohr | | | |
| $H_4$ | −2.19038 | −2.19038 | −2.19041 | −2.19038 | −1.92657 | −1.92658 | −1.94189 | −1.92658 |
| $H_6$ | −3.26674 | −3.26674 | −3.26674 | −3.26674 | −2.89051 | −2.89041 | −2.98320 | −2.89052 |
| $H_8$ | −4.34406 | −4.34507 | −4.34500 | −4.34508 | −3.83743 | −3.85455 | −3.88834 | −3.85458 |
| $H_{10}$ | −5.41879 | −5.42435 | −5.42419 | −5.42439 | −4.78592 | −4.81869 | −4.96708 | −4.81870 |
| $H_{12}$ | −6.49092 | −6.50415 | −6.50387 | −6.50423 | −5.65181 | −5.78283 | −5.98789 | −5.78285 |

## REFERENCES

(1) Sun, Q.; Berkelbach, T. C.; Blunt, N. S.; Booth, G. H.; Guo, S.; Li, Z.; Liu, J.; McClain, J. D.; Sayfutyarova, E. R.; Sharma, S.; Wouters, S.; Chan, G. K.-L., PySCF: the Python-based simulations of chemistry framework. *WIREs Comput. Mol. Sci.* **2018,** *8* (1), e1340.

(2) Sun, Q.; Zhang, X.; Banerjee, S.; Bao, P.; Barbry, M.; Blunt, N. S.; Bogdanov, N. A.; Booth, G. H.; Chen, J.; Cui, Z.-H.; Eriksen, J. J.; Gao, Y.; Guo, S.; Hermann, J.; Hermes, M. R.; Koh, K.; Koval, P.; Lehtola, S.; Li, Z.; Liu, J.; Mardirossian, N.; McClain, J. D.; Motta, M.; Mussard, B.; Pham, H. Q.; Pulkin, A.; Purwanto, W.; Robinson, P. J.; Ronca, E.; Sayfutyarova, E. R.; Scheurer, M.; Schurkus, H. F.; Smith, J. E. T.; Sun, C.; Sun, S.-N.; Upadhyay, S.; Wagner, L. K.; Wang, X.; White, A.; Whitfield, J. D.; Williamson, M. J.; Wouters, S.; Yang, J.; Yu, J. M.; Zhu, T.; Berkelbach, T. C.; Sharma, S.; Sokolov, A. Y.; Chan, G. K.-L., Recent developments in the PySCF program package. *J. Chem. Phys.* **2020,** *153* (2), 024109.

(3) Carleo, G.; Troyer, M., Solving the quantum many-body problem with artificial neural networks. *Science* **2017,** *355* (6325), 602-606.

(4) Choo, K.; Mezzacapo, A.; Carleo, G., Fermionic neural-network states for ab-initio electronic structure. *Nat. Commun.* **2020,** *11* (1), 2368.

(5) Yang, P.-J.; Sugiyama, M.; Tsuda, K.; Yanai, T., Artificial Neural Networks Applied as Molecular Wave Function Solvers. *J. Chem. Theory Comput.* **2020,** *16* (6), 3513-3529.

(6) Foster, J. M.; Boys, S. F., Canonical Configurational Interaction Procedure. *Rev. Mod. Phys.* **1960,** *32* (2), 300-302.

(7) Johnson, R., *Computational Chemistry Comparison and Benchmark Database* http://cccbdb.nist.gov/ DOI:10.18434/T47C7Z. 2022.

(8) Kim, S.; Thiessen, P. A.; Bolton, E. E.; Chen, J.; Fu, G.; Gindulyte, A.; Han, L.; He, J.; He, S.; Shoemaker, B. A.; Wang, J.; Yu, B.; Zhang, J.; Bryant, S. H., PubChem Substance and Compound databases. *Nucleic Acids Res.* **2016,** *44* (D1), D1202-13.